\documentclass[a4paper,usenatbib]{mnras}
\usepackage{aas_macros}
\usepackage{ae,aecompl}
\usepackage{graphicx}
\usepackage{amsmath}
\usepackage{amssymb}
\usepackage[dvipsnames]{xcolor}
\hypersetup{colorlinks=true,allcolors=teal}
\usepackage{ulem}

\title[Structural properties of LSB galaxies]{Structural properties of faint low surface brightness galaxies}

\author[Pahwa \& Saha]{
Isha Pahwa,$^{1}$\thanks{E-mail: ipahwa@iucaa.in}
and Kanak Saha$^{1}$\thanks{E-mail: kanak@iucaa.in}
\\
$^{1}$Inter-University Centre for Astronomy and Astrophysics, Ganeshkhind, Post Bag 4, Pune 411007, India\\
}

\date{draft}

\pagerange{\pageref{firstpage}--\pageref{lastpage}} \pubyear{2018}

\begin{document}

\label{firstpage}

\maketitle

\begin{abstract}
\noindent
We study the structural properties of Low Surface Brightness galaxies (LSB) using a sample of 263 galaxies observed by the Green Bank Telescope \citep{schneider+92}. We perform 2D decompositions of these galaxies in the SDSS $g$, $r$ and $i$ bands using the GALFIT software. Our decomposition reveals that about $60\%$ of these galaxies are bulgeless i.e., their light distributions are well modelled by pure exponential disks. The rest of the galaxies were fitted with two components: a Sersic bulge and an exponential disk. Most of these galaxies have bulge-to-total (B/T) ratio less than 0.1. However, of these 104 galaxies, $20\%$ have $B/T > 0.1$ i.e., hosting significant bulge component and they are more prominent amongst the fainter LSBs. According to $g - r$ colour criteria, most of the LSB galaxies in our sample are blue, with only 7 classified as red LSBs.  About $15\%$ of the LSB galaxies (including both blue and red) in our sample host stellar bars. The incidence of bars is more prominent in relatively massive blue LSB galaxies with very high gas fraction. These findings may provide important clues to the formation and evolution of LSB galaxies - in particular on the bar/bulge formation in faint LSB disks.
\end{abstract}

\begin{keywords}
galaxies: photometry --- galaxies: evolution --- galaxies: structure -- galaxies:spiral
\end{keywords}

\section{Introduction} 
\label{sec:intro}
The last two decades have witnessed a significant increase in the population of low surface brightness (hereafter, LSB) galaxies due to wide-field surveys such as the Sloan Digital Sky Survey~\citep[SDSS]{york+00}, as well as due to advancement in new techniques in observational astronomy allowing one to dig deeper in the sky \citep{Kniazevetal2004,zhong+08,rosenbaum+09,galaz+11,blanton+11,duc+15,TF16,greco+17}. This growing number of LSB population suggests that they might hold a significant fraction of baryon repositories in the local universe \citep{IB97,mcgaugh+95} although they occupy the faint end of the galaxy luminosity function \citep{impey+88,bothun+85}. In fact, the number density of LSB galaxies is comparable to that of the high surface brightness (hereafter, HSB) galaxies \citep{mcgaugh+95,OB+00,oneil+03}. But the formation and evolution of these LSB galaxies remained unclear and are thought to have taken a different route than the HSB galaxies.

There are several indications such as low star-formation rates \citep{hulst+93,schombert+11}, low metallicities \citep{mcgaugh+94,DV+98,kuzio+04}, sparse H-$\alpha$ emission~\citep{pickering+97,huang+14} suggesting that LSB galaxies are under-evolved system compared to their HSB counterparts \citep{bothun+97}. In general, these LSB galaxies are known to have high neutral hydrogen gas to stellar mass ratios, associated with nearly non-detection or little CO molecules \citep{Oneil+03a,honey+18} - indicating their inefficiency in converting the gas to stars. Although the exact reason remains to be understood, the combination of low surface density and dark matter dominance at all radii, as derived from the observed rotation curve \citep{deBM96, deblok+01} makes sure that the disk instabilities are unlikely to set in. In fact, using analytical and numerical simulations which include gas and cold dark matter components, it has been shown that the realistic models of LSB galaxies are stable against local and global instabilities \citep{Mihos+97, Mayer-Wadsley2004, GJ14}. In other words, LSB galaxies would be unlikely to host strong bars and spiral features as reflected in earlier observations by \cite{mcgaugh+94, Impeyetal1996} who found bar fraction to be only a few percent. But several recent studies seem to indicate a growing bar fraction from $\sim 8\%$ \citep{Honeyetal2016} to about $20\%$ in \cite{SG17} as larger samples of LSB galaxies are being analyzed. If such a trend continues, one needs to rethink about the evolution of normal LSB galaxies. 

Previous studies that have explored in detail the properties of LSB galaxies are mostly late-type disk dominated \citep{blok+95, mcgaugh+94} or Malin-type giant LSBs \citep{Sprayberry+95,pickering+97}. Over the last ten years, this notion seems to be changing as HST revealed an insightful picture of Malin1 - the very inner part has a bar and a bulge \citep{Barth2007} - like in a normal HSB galaxy. A similar conclusion was derived by \cite{Lellietal2010} who found Malin1 to have a normal HSB like inner disk. A number of other LSBs are reported to host bulges whose stellar populations, colours and gas kinematics are remarkably similar to those hosted by HSB galaxies \citep{Beijersbergenetal1999, Galazetal2006, Pizzellaetal2008,Morellietal2012} - indicating the LSBs might be having a parallel formation sequence to the HSB galaxies. This has prompted us to investigate the two component bulge-disk decomposition of a sample of LSB galaxies that remained relatively unexplored in the literature.

In the present work, we choose a sample of LSB galaxies observed by the Green Bank Telescope (GBT) as given in  \cite{schneider+92} to study the structural properties of LSB galaxies. We perform two component bulge-disk decomposition of 294 LSB galaxies using GALFIT \citep{galfit+02,galfit+10}. Further based on visual inspection, the sample is divided into two groups - barred and unbarred and their properties are discussed in detail.

This paper is organised as follows. In Section~\ref{sec:sample_method}, we present our galaxy sample. We also discuss the 2D decomposition steps and the method to obtain the structural parameters in this section. The results are presented in Section~\ref{sec:results}. Lastly, we discuss and summarize the conclusions in Section~\ref{sec:discussion}.

\section{Sample selection and method}
\label{sec:sample_method}

\subsection{The GBT/SDSS Sample}
\label{sec:sample}
The Uppsala General Catalog of galaxies \citep[UGC]{nilson1973} consists of a list of dwarf and LSB galaxies which was assembled by Nilson in 1973. This catalog covers all northern galaxies ($\delta \ge -2^{\circ} 30'$) visible on the Palomar Sky Survey with a blue diameter larger than $1'$. In the nineties, a group led by S. E. Schnieder used this sample to map the neutral hydrogen of dwarf and other low surface brightness (LSB) galaxies such as  irregular, Sd-m etc. This selection of galaxies by S. E. Schneider brings an incompleteness of $\sim$14\% in the sample. His group took neutral hydrogen observations in two steps. First, they used Arecibo telescope for the galaxies in the declination range $-2^{\circ} \le \delta \le 38^{\circ}$. They reported 762 dwarf or LSB galaxies with this telescope \citep{schneider+90}. In the second step, they used Green Bank telescope (GBT) for galaxies in the declination range, $\delta \ge 38^{\circ}$ along with a number of galaxies farther south for flux comparisons with Arecibo observations and to search for extended halos, totalling 633 galaxies \citep{schneider+92}. We choose their sample of 633 galaxies to analyse the detailed structural properties. The sample completeness is discussed in \cite{schneider+90,schneider+92,thuan+91,DT+96} who have also studied the spatial clustering and their relationship to bright galaxies.

The GBT sample of 633 galaxies is retrieved from NASA/IPAC EXTRAGALACTIC DATABASE (NED) which provides the $ra$, $dec$, redshift $z$ etc. \citep{schneider+92}. We cross-match this sample with SDSS \citep{york+00} data release 13 \citep[DR13]{DR13} and find $354$ galaxies for which spectroscopic redshifts are available. Finally, we have visually inspected images in $r$ and $g$ bands of each of these galaxies to remove from the sample those galaxies which are drastically affected by a merger or a companion and also those galaxies where the presence of a bright star is affecting the analysis. These are a total of 19 in number.  We have also removed 35 edge-on galaxies from sample. In addition to them, we have also rejected 6 galaxies having star-forming clumps in them. This has further reduced our sample to $294$ galaxies that are common between GBT and SDSS. These galaxies are in the redshift range of 0.001-0.037.

\begin{figure*}
\centering
\includegraphics[width=0.7\textwidth,height=0.6\textwidth]{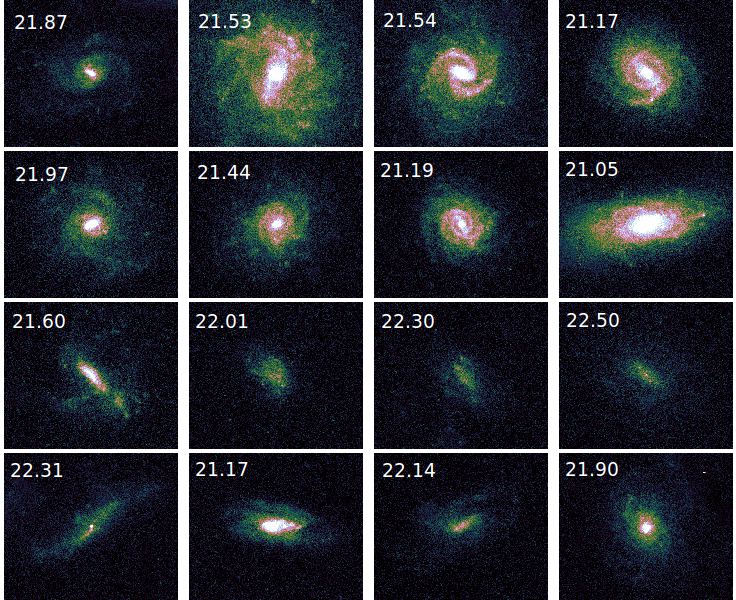}
\caption{A few representative galaxies of different morphologies in our sample in $r$-band. The disk central surface brightness (in units of mag/arcsec$^2$) is indicated on the top of each galaxy. The colour scale is same for all images.}
\label{fig:lsb_mix}
\end{figure*}

In order to compare our LSB sample with other disk galaxies on the colour magnitude plane, we use the \cite{simard+11} catalog which is based on SDSS DR7. In particular, we use their bulge-disk decomposition parameters where bulges are fitted with free Sersic indices. We extract galaxies in the same redshift range as of our sample. This final sample consists of 70810 galaxies for which we have $r$-band absolute magnitudes, redshifts and bulge-to-disk ratios. We call this sample as ``all galaxy sample''.

\begin{figure*}
\centering
\includegraphics[width=0.7\textwidth,height=0.6\textwidth]{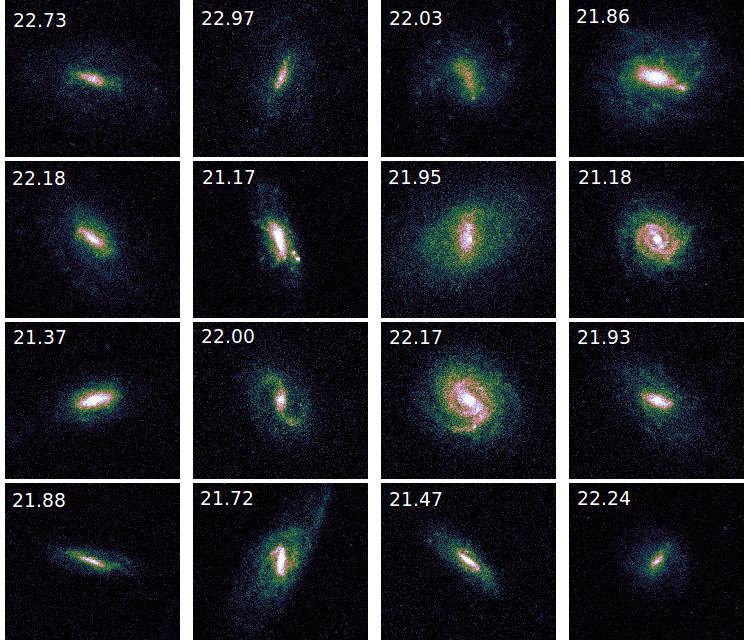}
\caption{Same as in Figure~\ref{fig:lsb_mix} but here we have chosen to show only barred galaxies in our sample.}
\label{fig:lsb_bar}
\end{figure*}

\subsection{Multi-band Bulge-Disk Decomposition}

The images of $294$ galaxies are drawn from the SDSS $g$ (green), $r$ (red) and $i$ (near-infrared) filters with effective central wavelengths being 4770, 6231 and 7625 A$^\circ$. The SDSS Science Archive Server provides the survey images for these galaxies, called ``corrected frames'' which are identified by a unique name and it's a combination of  run number, the camera column and the frame sequence number. Figure~\ref{fig:lsb_mix} shows the $r$-band images of some of these galaxies with mix morphologies e.g., spirals, irregulars from our sample. These images are calibrated in nanomaggies per pixel, and have a sky-subtraction applied. From each frame, a 2-4 arcmin cutout from the central coordinate of the galaxy is extracted depending upon the galaxy's redshift. It ensures that an average galaxy size covers at least 50$\%$ of the total area.

Before we proceed to discuss the bulge-disk decomposition, we decontaminate these 294 (in each of the three bands) galaxies to remove the surrounding sources around the target galaxy in the cutout. For removing these unwanted sources, each galaxy image cutout is considered  separately and the pixel values of each unwanted source are replaced with the average value of the pixels surrounding that source. This is done using the Image Reduction and Analysis Facility (IRAF) IMEDIT task, which creates a circular annulus of a chosen radius around the central coordinates of the selected source and replaces the pixels in this circle with the background values.

Once the cutouts are cleaned, we perform two-dimensional bulge-disk decomposition on each galaxy image using the GALFIT (version 3.0.5) software \citep{galfit+02,galfit+10}. This software models the light distribution using the analytic functions, known as parametric fitting which adjusts the parameters in the analytic functions to try and match with the shape and profile of galaxies. For more details about the fitting algorithm and the usage of GALFIT, the reader is advised to go through \cite{galfit+02,galfit+10}.    

\begin{figure}
\includegraphics[width=0.5\textwidth,height=0.45\textwidth]{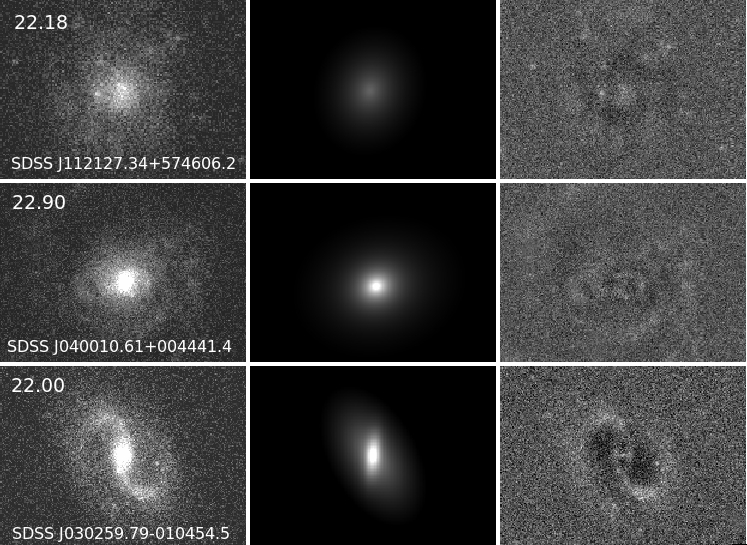}
\caption{Three examples of bulge-disk decompositions using GALFIT. The first column shows the $r$-band observation images of the selected galaxies, the second column shows the GALFIT model images and the third column shows the residual images which are basically the subtraction of model images from the observation images. The three selected galaxies are \textit{typical} examples of irregular, bulge and bar galaxies. The disk central surface brightness and SDSS IDs are indicated on the top of each galaxy.}
\label{fig:residual}
\end{figure}

We start with by fitting two components, i.e., bulge and disk profiles to the light distribution of each galaxy in $r$-band. We adopt the standard Sersic and the exponential profiles for the bulge and disk components of the galaxy, respectively \citep{sersic68,freeman70}. These are discussed as follows -

\begin{itemize}
\item The Sersic profile has the following functional form:
\begin{equation}
I(R) = I_e \, \mathrm{exp}\left\{-b_n[(R/R_e)^{1/n} -1]\right\},
\end{equation}
where $I(R)$ is the pixel surface brightness at a radius $R$ from the centre of a galaxy, the parameter $R_e$ is known as the effective radius such that half of the total flux is within $R_e$ and $I_e$ is the pixel effective surface brightness at the effective radius $R_e$ and the parameter $n$ is the sersic index that controls the shape of the light profile. The dependent-variable $b_n$ is coupled to $n$ and is given as $b_n = 1.9992n-0.3271$ \citep{GD05}, thus it is not a free parameter. 

\item The exponential profile for the disk is given by:

\begin{equation}
I(R) = I_0 \, \mathrm{exp}\left(-\frac{R}{R_s}\right),
\end{equation}
where $I_0$ is the central surface brightness, $R_s$ is the scale length of the stellar disk. 
\end{itemize}
The full profile is the addition of the Sersic and the exponential profiles. Bars in the sample are included in the bulge light and are not dealt separately.  
To run the GALFIT successfully, we need to provide the point spread function (PSF). To generate PSF images for SDSS observations, a Gaussian profile with a given Full Width at Half Maximum (FWHM) of the surface brightness distribution is fit with GALFIT. The FWHMs for SDSS observations are obtained from Science Archive Server. The background image (also known as $\sigma$ image) is generated internally.

In addition, the GALFIT software also requires initial guesses of parameters of bulge and disk profiles which we choose from SDSS such as $ra$, $dec$, $PetroMag\_r$, $PetroMag\_g$, $PetroMag\_i$, $deVAB\_r$, $expAB\_r$, $deVPhi\_r$, $expPhi\_r$. After setting the initial parameters, we run the GALFIT for all 294 galaxies in $r$-band. The output of the GALFIT fitting returns the final model of the galaxy and the residual image which is formed by the subtraction of the final model from the original image. All the residuals are visually inspected to see whether the final model obtained is a good fit to the original image or not. On the basis of residual inspection and bad reduced $\chi^2$ values, we do not include 31 galaxies in our study. As mentioned earlier in this section, we have rejected few galaxies having star-forming clumps around the centre. This was also done based on the visual inspection. As GALFIT was not able to fit them properly (reduced $\chi^2$ was not good), the residuals of these type of galaxies had left-over bright sub-components which showed that these galaxies were having star-forming clumps. We thus do not include them in our study.  Our final sample consists of 263 galaxies.\footnote{The online supplementary material consists of GALFIT output for these 263 galaxies in our sample. For all these galaxies, reduced $\chi^2$ is around one. The image and GALFIT fits are available on request.} All the subsequent analysis and results presented in this paper are based on this specific sample. Based on the visual inspection in $r$ and $g$ bands, we have found 43 bars out of 294 galaxies i.e., $\sim 15\%$ are barred LSBs in our sample. In Figure~\ref{fig:lsb_bar}, we show images of some representative barred LSBs in our sample. A discussion on the barred LSBs is presented in section~\ref{sec:bar-HIgas}.

The output of the GALFIT consists of three images -
\begin{enumerate}
\item The postage stamp sized region of the input image.
\item The final model of the galaxy in that region.
\item The residual image which is formed by subtracting the final model from the first image.
\end{enumerate}
In Figure~\ref{fig:residual}, we show the three examples of bulge-disk decompositions which we have obtained using GALFIT software. The three selected galaxies are \textit{typical} examples of irregular, bulge and bar galaxies as shown in first, second and third row. The first column shows the $r$-band observation images of these selected galaxies, the second column shows the GALFIT model images and the third column shows the residuals images.

After fitting all the galaxies in our sample in $r$-band, we follow the same procedure to fit the galaxies in $g$ and $i$ bands. We make the cutouts accordingly and decontaminate them as described previously. We generate PSF using the same procedure. Now, we take the advantage of fitting the galaxies in $r$-band. We apply the results of $r$-band decompositions to the $g$ and $i$ bands decompositions by fixing all the parameters except for the positions of centres and the bulge and disk magnitudes of the galaxies. This technique, i.e., using results of one band into others as initial conditions, is known as ``simultaneous fitting'' technique and has been used extensively in literature \citep{simard+11,lackner+12,meert+15,kim+16}. 

\subsection{Structural parameters}
The two-component decomposition by GALFIT provides us the basic photometric parameters of the bulge and disk components for each galaxy in our sample. These parameters are further used to derive quantities such as central surface brightness, colour of the galaxy etc. as explained below. 
We convert the model magnitude to AB magnitude system  using standard relation \citep{Oke1974}. All the magnitudes are then corrected for Galactic Extinction. Hereafter, all the magnitudes mentioned in this paper correspond to the AB magnitude system unless otherwise specifically mentioned. 

The logarithmic central surface brightness for the disk component ($\mu_{0,\mathrm{disk}}$) can be calculated as
\begin{equation}
\mu_{0,\mathrm{disk}} = m_{\mathrm{disk}} + 2.5 \log_{10}(2\pi R_s^2),
\end{equation}
where $m_{\mathrm{disk}}$ refers to the apparent magnitude of the disk component of the galaxy. The total central surface brightness of the galaxy is given by
\begin{equation}
I_{\mathrm{total}} = I_0 + I_e \, \mathrm{e}^{b_n}
\end{equation}
and the logarithmic total central surface brightness can be written as
\begin{equation}
\mu_{0,\mathrm{total}} =  \mathrm{zpt} -2.5 \log_{10} I_{\mathrm{total}}.
\end{equation}

The central surface brightness is then corrected for the inclination and the cosmological dimming effects by using the following equation \citep{zhong+08}
\begin{equation}
\mu_{0,\mathrm{corrected}} = \mu_0 + 2.5 \log_{10} (b/a) - 10 \log_{10} (1+z).
\end{equation}
Following on, we will always use ``corrected'' central surface brightness everywhere and thus we will drop the subscript ``corrected'' from the definition of $\mu_0$.  

After estimating $\mu_0$ in $r$ and $g$ bands, we calculate the approximate $B$-band central surface brightness by using the following transformation equation~\citep{smith+02,SG17}:
\begin{equation}
\mu_0(B) = \mu_0(g) + 0.47 (\mu_0(g)-\mu_0(r)) + 0.17.
\end{equation} 
This will be used to classify our sample galaxies as LSBs as per the criterion given in $B$-band. One needs to be bit cautious here as it is an approximate transformation equation. It can aid to the uncertainty in the results.

The absolute magnitude, $M$ is corrected against dust extinction using $K$-correction as follows:
\begin{equation}
M = m - 5 \log_{10}\left(\frac{D}{10 \mathrm{pc}}\right) - K,
\end{equation} 
where $m$ is the apparent magnitude of the galaxy, $D$ is the distance to the galaxy in pc and $K$ is the $K$-correction term that accounts for the difference to transfer from observed band to the rest-frame band. The correction term depends on the spectral energy distribution and hence on the redshift of the object \citep{hogg+02}. The $K$-correction assumes a simple relation for a power-law continuum
\begin{equation}
K = -2.5 (1+\alpha_{\nu}) \log_{10}(1+z),
\end{equation}
where $\alpha_{\nu}$ is the slope of the continuum and has a canonical value of -0.5~\citep{SG83,boyle+88}.

\begin{figure}
\includegraphics[width=0.45\textwidth]{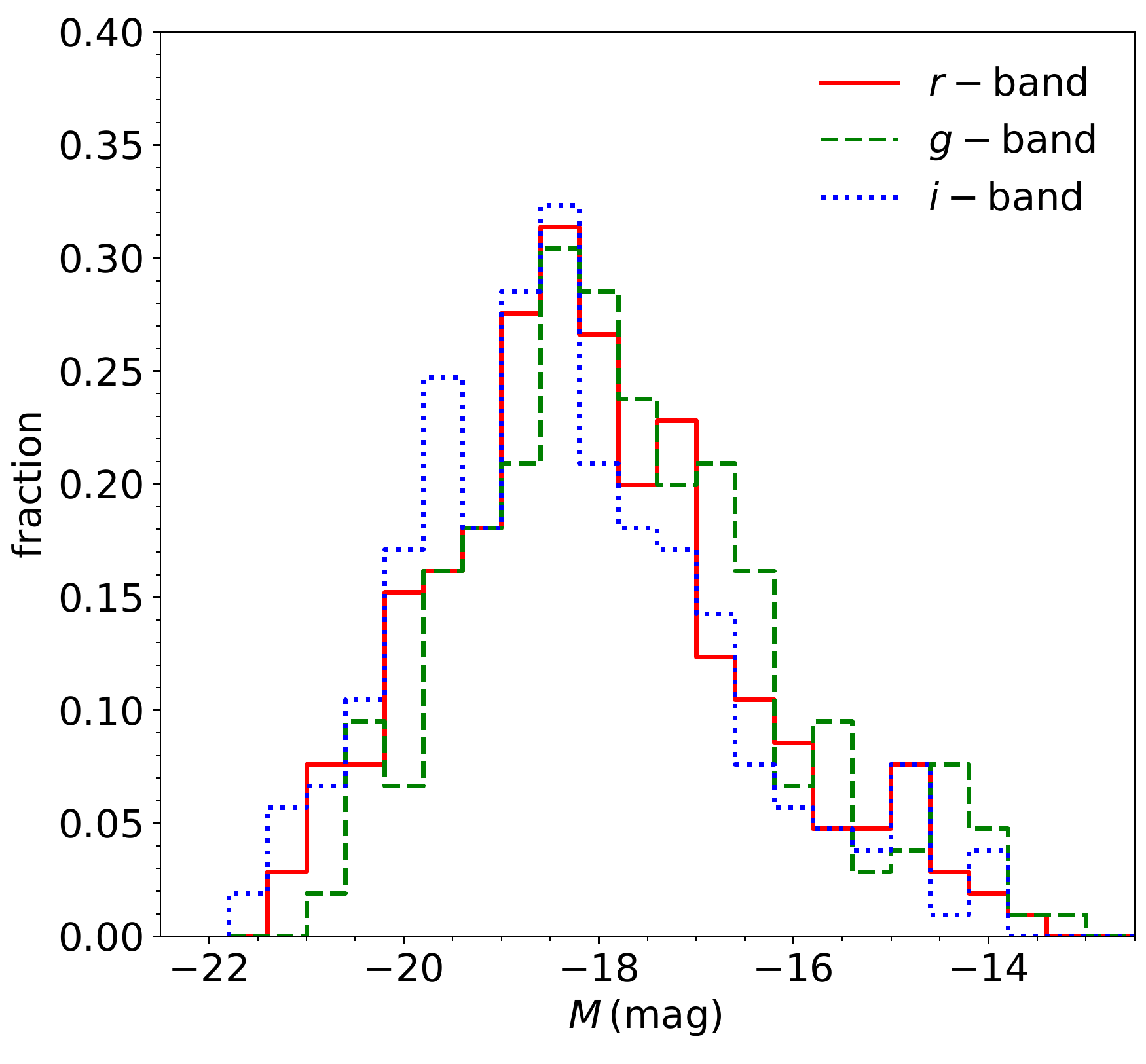}
\caption{Histogram of the absolute magnitudes of the galaxies in our sample in three bands, namely $r$, $g$ and $i$ as shown by red (solid), green (dashed) and blue (dotted) lines respectively. The absolute magnitude here corresponds to the combined magnitude of the bulge and disk components of the galaxies.} 
\label{fig:hist_abs_mag}
\end{figure}

Figure~\ref{fig:hist_abs_mag} depicts the distribution of the absolute magnitudes of our sample galaxies in $r$, $g$ and $i$ bands, where the absolute magnitude refers to the total model magnitude obtained from our GALFIT analysis. Our sample galaxies cover a wide range of magnitudes, while some are as bright as $M_{r} \sim -21$; the dominant population of the galaxies are on the fainter side ($M_{r} \ge -19$) going upto about -14. The medians of distributions of the absolute magnitudes of our sample galaxies in $r$, $g$ and $i$ bands lie at -18.07, -17.71 and -18.26. 
It seems from this figure that these galaxies are overall more luminous in the $i$-band (i.e., near-infrared) by about 0.19-0.55 mag than in $r$ and $g$ bands (optical bands) which indicates the ageing of the existing stars in these galaxies. 

\section{Distribution of central surface brightness}
\label{sec:results}
The classification of a galaxy into a HSB or LSB is based on its disk central surface brightness ($\mu_{0,\mathrm{disk}}$). A disk galaxy is called HSB if its $B$-band disk central surface brightness peaks at 21.56 $\mathrm{mag/arcsec}^2$~\citep{freeman70}. Disk galaxies whose disk central surface brightness is about $1$~mag fainter, are termed as LSBs. According to this, we define LSB galaxies with $\mu_{0,\mathrm{disk}} \ge 22.5 ~\mathrm{mag/arcsec}^2 $ in the B-band  \citep{mcgaugh96,rosenbaum+09}.

A similar criterion has been adopted to classify a disk galaxy as an LSB  in the $r$-band, i.e.,
$ \mu_{0,\mathrm{disk}}(r) = 21 \, \mathrm{mag/arcsec}^2$~\citep{courteau96,brown+01,adami+06} and is $\sim 1.1\sigma$ away from the mean value of HSB galaxies as reported by \cite{courteau96}. 

\begin{figure}
\includegraphics[width=0.45\textwidth]{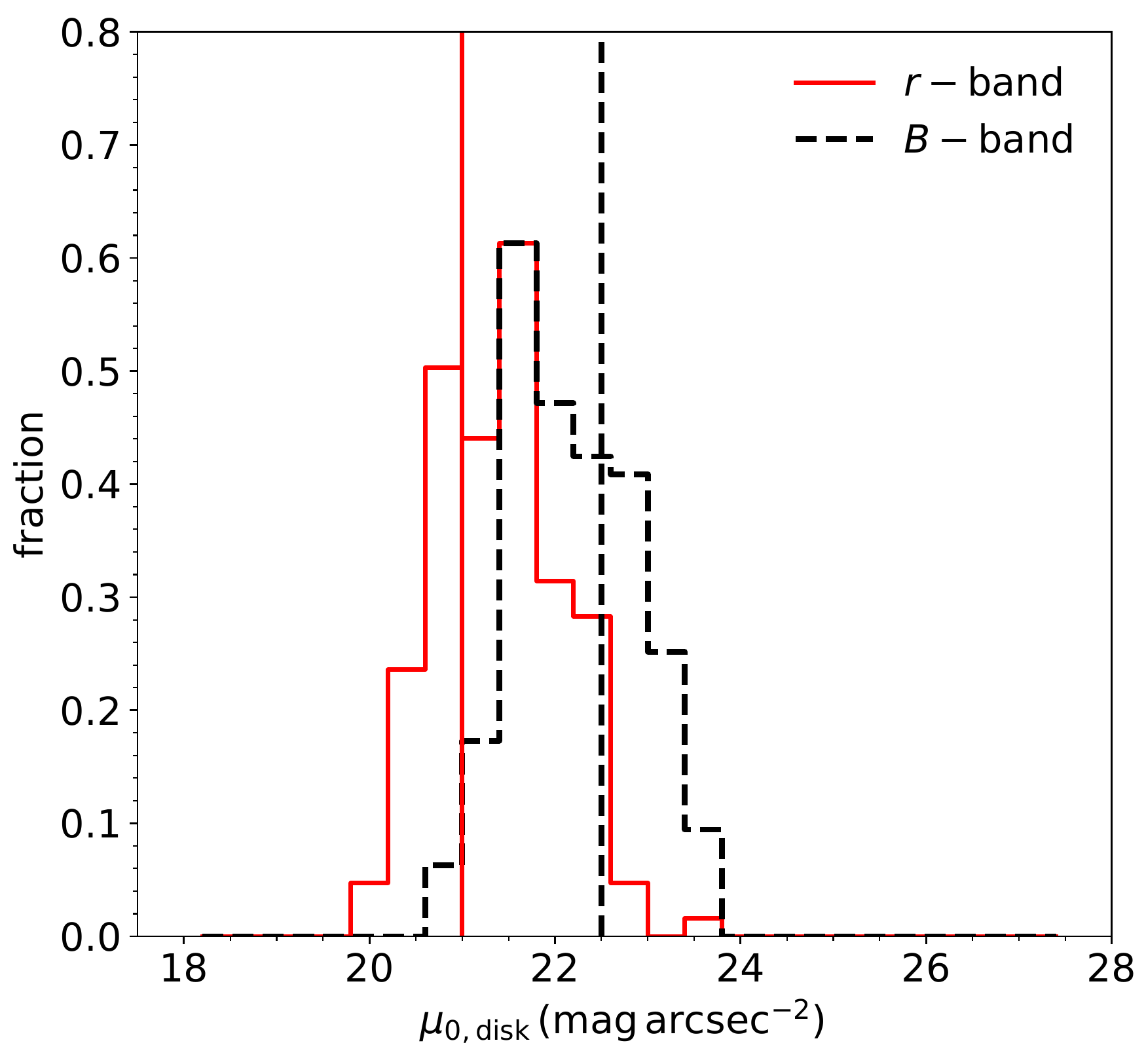}
\caption{{\it Bulgeless LSB galaxies:} histogram of the disk central surface brightness ($\mu_0$) of the galaxies in our sample in $r$ and $B$ bands as shown by red (solid) and black (dash-dotted) lines respectively. We have only shown those galaxies where \textit{only disk} component fitting has been done using GALFIT software. The vertical solid (red) and dashed (black) lines correspond to the thresholds of $\mu_0$ being 21 and 22.5 $\mathrm{mag \,arc/sec}^2$ in $r$ and $B$ bands, i.e., galaxies lying to the right of these lines in the distribution are LSB galaxies.} 
\label{fig:hist_disk_SB}
\end{figure}

\begin{figure}
\includegraphics[width=0.45\textwidth]{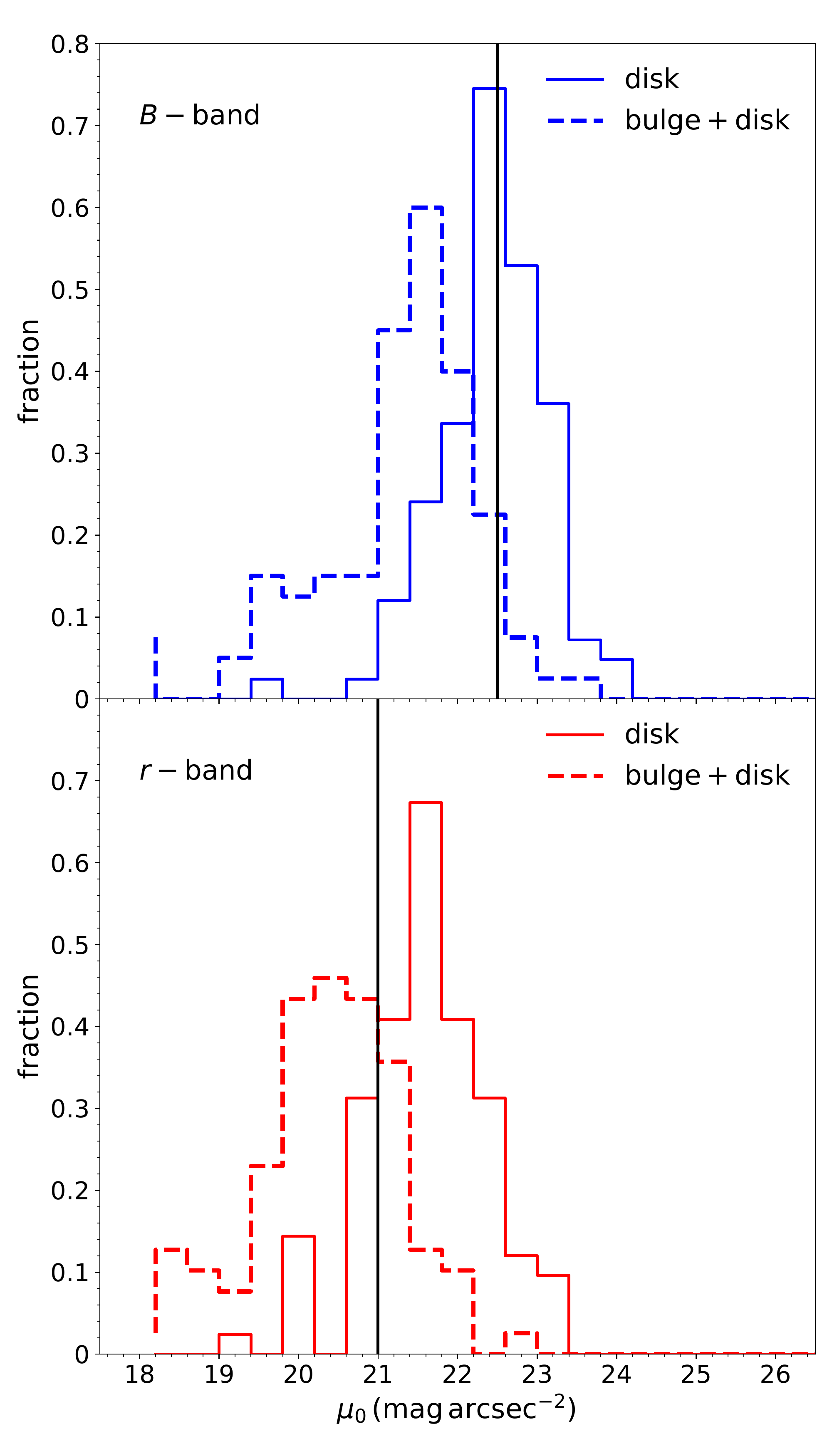}
\caption{Same as in Figure~\ref{fig:hist_disk_SB} but for the {\it LSB galaxies with bulges}. The disk and total (bulge+disk) central surface brightness ($\mu_0$) of the galaxies in our sample in $r$ (lower panel) and $B$ (upper panel) bands are shown as solid and dashed lines.}
\label{fig:hist_SB}
\end{figure}

In the rest of this paper, we follow this $r$-band criterion to select LSB from our sample, i.e., a galaxy is termed as LSB whose $\mu_{0,\mathrm{disk}} \ge 21 \, \mathrm{mag/arcsec}^2$ in the $r$-band.  \cite{schneider+92} have already filtered out the LSB galaxies from the UGC catalog, however, we re-confirm this by plotting their distribution of disk central surface brightness in Figures~\ref{fig:hist_disk_SB} and \ref{fig:hist_SB}. 

\begin{figure}
\includegraphics[width=0.45\textwidth]{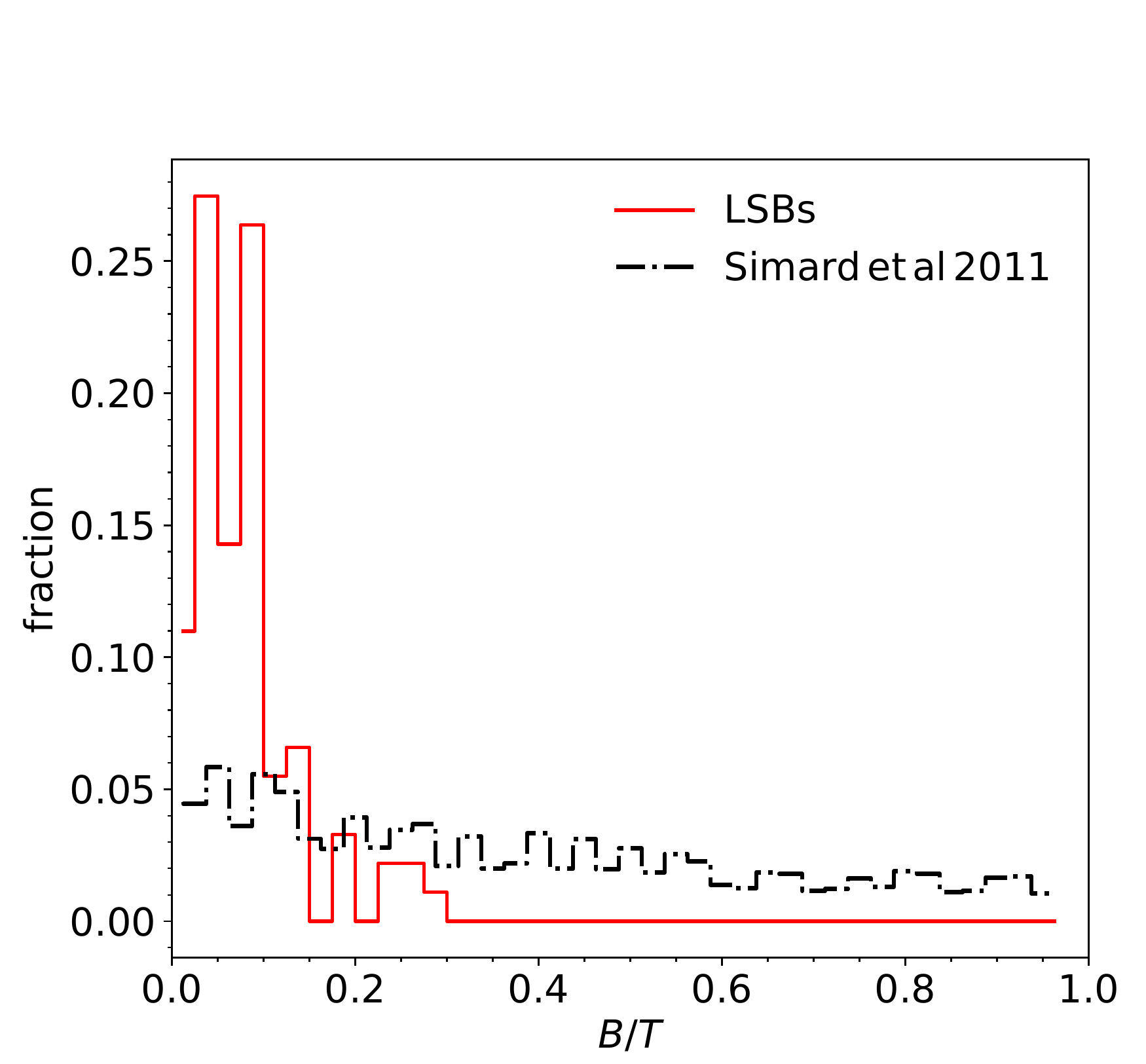}
\caption{Bulge-to-total (B/T) ratios for our LSB sample and  {Simard et al 2011} sample. Around 80\% of galaxies in our sample have $B/T$ less than 0.1.}
\label{fig:BT_totmag}
\end{figure}

\subsection{Bulgeless LSB galaxies}
\label{sec:bulgeless}
About $60\%$ (i.e., 159) galaxies in our sample are fitted well with a single exponential disk model without any conspicuous bulge component. Visual inspection of their morphologies suggests that these LSBs are either smooth disks or irregulars without any brighter central part. The top panel of Figure~\ref{fig:residual} shows a typical example of these bulgeless LSB galaxies and their 2D decomposition in the $r$-band. Figure~\ref{fig:hist_disk_SB} shows their distribution of central surface brightness. The vertical solid (red) and dashed (black) lines correspond to the thresholds of $\mu_0$ being 21 and 22.5  $\mathrm{mag \,/arcsec}^2$ in $r$ and $B$ bands respectively. The galaxies lying on the right of these lines are classified as LSBs, and $81\%$ galaxies in our sample are LSBs as per $r$-band criteria. The median of $r$-band distribution lies at 21.71 $\mathrm{mag \,/arcsec}^2$ and that of $B$-band lies at 22.51 $\mathrm{mag \,/arcsec}^2$. The medians of $\mu_0$ distributions for $g$ and $i$ bands lie at 22.11 and 21.56 $\mathrm{mag \,/arcsec}^2$ respectively. It is worth mentioning here that the  population of bulgeless galaxies in the local universe are not only limited to HSB but also exist in the LSB regime. According to \cite{Kautsch2009}, the fraction of bulgeless galaxies in edge-on projection is about $15\%$.  \cite{FY11} also quotes $\sim 35\%$  of their sample as bulgeless disk galaxies.  
Finding these large-sized disks with no bulges in the local universe therefore challenges our understanding of galaxy formation in the hierarchical framework \citep{WhiteRees1978}. Since LSBs are generally found in isolated or rather less dense environments, they might have possibly avoided mergers otherwise their disks would have some amount of bulge component and pure exponential profile would be hard to maintain \citep{hopkins+10,stinson+13,naab+14}. Even if an LSB galaxy avoided mergers, internally driven secular evolution (due to non-axisymmetric features in the early phase of evolution) is inevitable and would produce central concentration or a bulge component \citep{LK72,KK04,kormendy15}. 

Recently, several cosmological hydrodynamical simulations have succeeded in producing reasonable disk galaxies; but maintaining them as bulgeless over longer period has remained as a harder task, most galaxies end up with $B/T \sim 0.3$, if not more \citep{Governato+10,Agertz+11,Guedes+11,Brook+12,Marinacci+14}. In a recent study, a similar fraction of bulgeless pure disks is found at $z\sim1$ \citep[by][]{SachdevaSaha2016} but these were rather massive, bright galaxies. The progenitors of these bulgeless LSBs would be interesting to look for in order to understand their early assembly and evolution since then.

\begin{figure*}
\includegraphics[width=\textwidth]{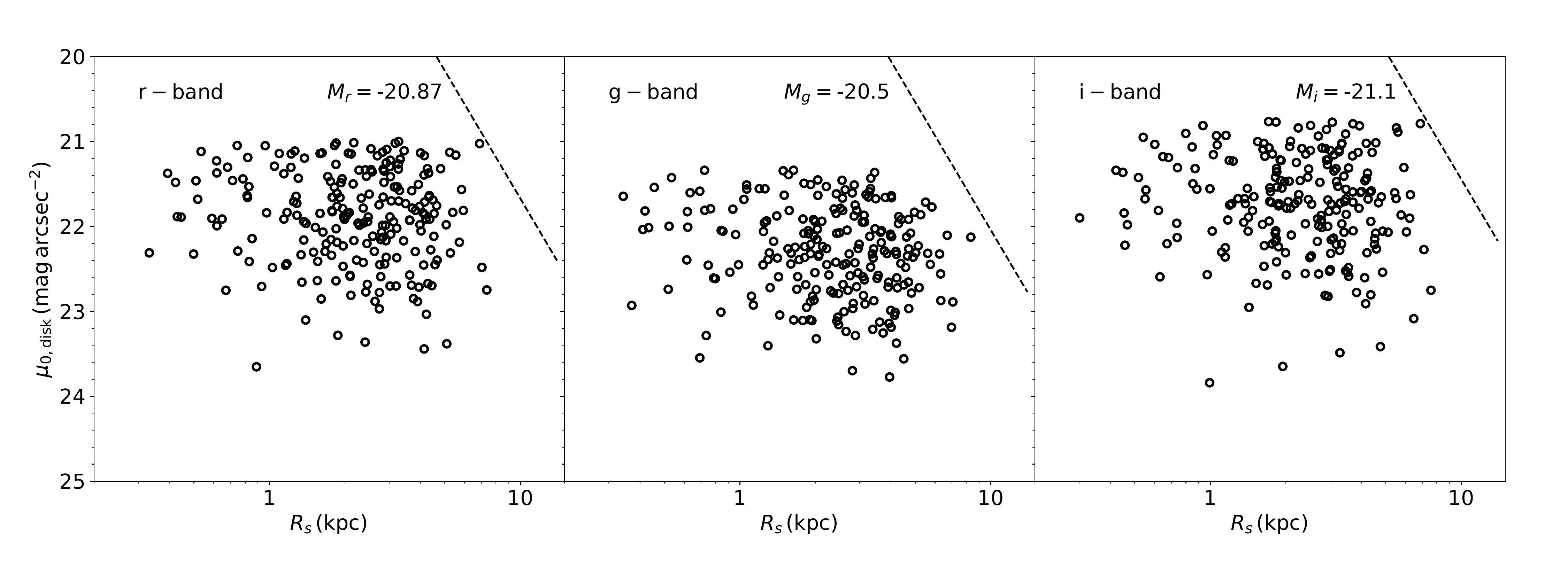}
\caption{The central surface brightness $\mu_{0,\mathrm{disk}}$ versus scale-length $R_s$ of the LSB galaxies by considering disk \textit{only} components in $r$-band (left panel), $g$-band (middle panel) and $i$-band (right panel). The dashed lines correspond to the exponential disks with constant absolute luminosity  of indicated magnitudes in each panel. The equality lines of other magnitudes lie parallel to the dashed line.}
\label{fig:disk_scale_length}
\end{figure*}

\begin{figure*}
\centering
\includegraphics[width=0.7\textwidth,height=0.7\textwidth]{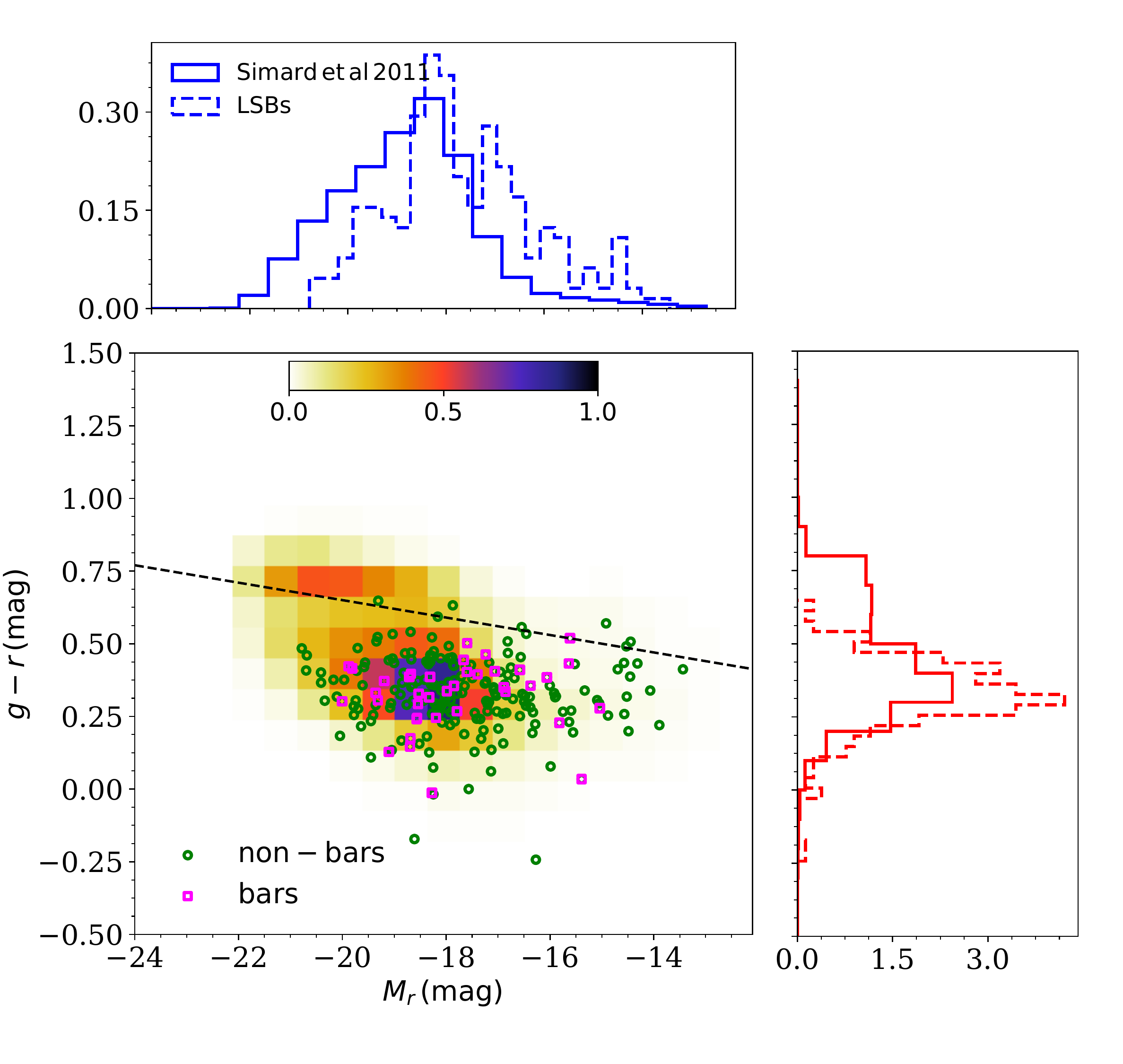}
\caption{The colour-magnitude diagram (CMR) of the LSB galaxies. This uses $g-r$ colours and the $r$-band absolute magnitudes of our LSB sample galaxies and all galaxy (Simard et al 2011) sample. In the middle panel, we show bars in our sample as magenta squares and all other galaxies as green circles. The density contours represent the all galaxy (Simard et al 2011) sample. This is over-plotted in order to compare our galaxies with the all galaxy sample. The black dashed line divides the sample into red and blue based on colour-cut. The top panel shows the distributions of $r$-band absolute magnitudes of our LSB sample galaxies (dashed line) and all-galaxy sample galaxies (solid line) whereas the right-hand side panel shows the distributions of $g-r$ colours where lines denote the same samples as that of top panel.}
\label{fig:colour_diskmag}
\end{figure*}

\begin{figure}
\includegraphics[width=0.45\textwidth]{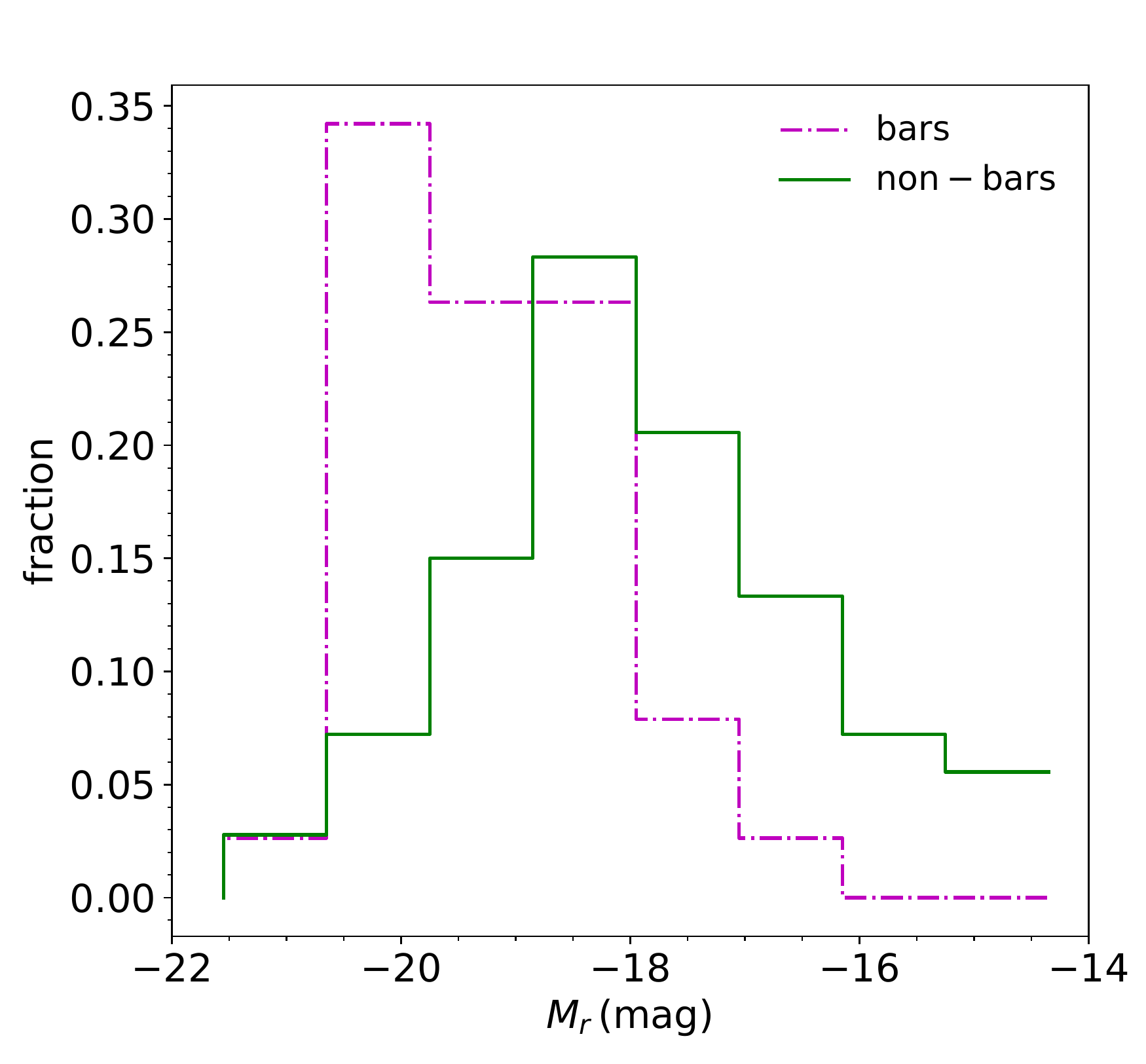}
\caption{This figure shows the histogram of absolute $r$- band magnitude of barred and unbarred galaxies. The barred galaxies are systematically brighter than the unbarred galaxies.}
\label{fig:bar_lum}
\end{figure}

\subsection{LSB galaxies with bulges}
\label{sec:LSBbulges}
For the rest of the galaxies ($\sim 40\%$ i.e., 104) in our sample, we were able to model the observed light distribution with two components, namely a bulge and disk (as explained above). In Figure~\ref{fig:hist_SB}, we show the distribution of the central surface brightness derived from the modelling. Due to the presence of the bulge component, the central surface brightness of these galaxies are brighter. If we use the total central surface brightness (as shown by the dashed line in Figure~\ref{fig:hist_SB}), the fraction of qualified LSBs obviously reduces.  A comparison of disk $\mu_0$ with the threshold dictates that $\sim$ 84 \% galaxies in this sample (where two component fitting has been done)  are LSBs. However, this fraction reduces to 61\% when the central surface brightness is governed by the full profile of both bulge and disk components. 

Figure~\ref{fig:BT_totmag} shows the bulge-to-total ratio ($B/T$) for our LSB sample galaxies and the all-galaxy population of \cite{simard+11} in the $r$-band. About $20\%$ of galaxies in our sample have $B/T$ more than 0.1. Although majority of the LSBs are with $B/T < 0.1$, it is surprising to see some of these are having significant bulge light.

\begin{figure*}
\centering
\includegraphics[width=\textwidth]{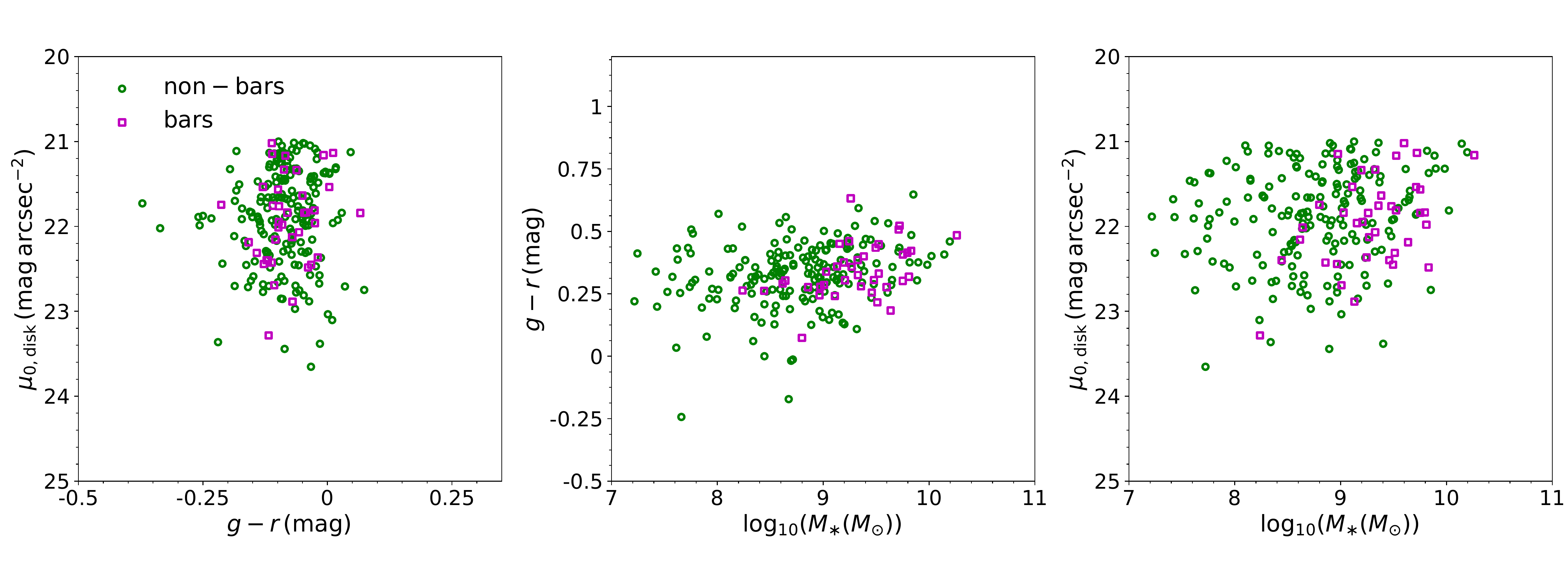}
\caption{This figure shows the relation between different structural properties of LSB galaxies. The left and middle panel show the scatter plot between the $r$-band $\mu_0$ and the stellar masses of the galaxies with the $g-r$ colours whereas the right panel shows the correlation between $r$-band disk $\mu_0$ and  stellar masses  of galaxies.}
\label{fig:struc_lsb}
\end{figure*}

\subsection{Disk scale-length vs central surface brightness}
\label{sec:scalelength}
The exponential (disk) scale length ($R_s$) of a galaxy is a fundamental structural parameter to model its dynamics and to constrain its formation mechanisms. Since LSB galaxies are believed to have different formation mechanisms than HSB galaxies \citep{bothun+97}, it motivates us to examine the correlation (if any) of the central surface brightness and the disk scale length. Figure~\ref{fig:disk_scale_length} shows the $\mu_{0,\mathrm{disk}}$-$R_s$ relation for all the LSB galaxies in our sample in $g$, $r$ and $i$ bands. The dashed line in each panel shows the $\mu_{0,\mathrm{disk}}-R_s$ relation for the constant disk luminosity and sets the upper limit in $\mu_{0,\mathrm{disk}}-R_s$ plane as there can not be galaxies with the large scale-lengths and high central surface brightness. The constant disk luminosity lines to constrain the degeneracy of $\mu_{0,\mathrm{disk}}$ and scale-length was first used by \cite{grosbol85}. Figure~\ref{fig:disk_scale_length} shows that there is no apparent correlation between $\mu_{0,\mathrm{disk}}$ and $R_s$ for our sample LSB galaxies  and this result is consistent with a number of other studies \citep{impey+88,davies+88,irwin+90,mcgaugh+94,zhong+08}. However, there is a tendency for the galaxies with the largest $R_s$ to be clustered in the low surface density regime and thus with low  $\mu_{0,\mathrm{disk}}$. In other words, the fainter LSB galaxies are associated with larger scale lengths. This has also been seen by \cite{mcgaugh+95b,dejong96,fathi10}. Apparently, most of these galaxies with large scale-lengths are quite isolated. This isolation helps them surviving against tidal interactions and mergers activities. It also seems that these large objects must have proceeded by the gradual accretion without any violent star formation or disruption by neighbouring systems~\citep{bothun+93,mcgaugh+94} so that they could collapse into a single object with low surface density rather than fragmenting into smaller objects. 

\subsection{Colour-Magnitude Relation}
\label{sec:CMR}
The colour-magnitude relation (CMR) is a powerful tool to understand the underlying stellar population and evolution of galaxies \citep{holmberg58,roberts+94}. 

Figure~\ref{fig:colour_diskmag} depicts the CMR for our LSB sample (marked as green circle and magenta square symbols) over plotted on the ``all-galaxy'' sample. The top panel shows the distribution of the $r$-band absolute magnitudes for our LSB sample (dashed line) and all-galaxy sample (solid line). The right-hand side panel shows the distributions of $g - r$ colour for our LSBs and all-galaxy sample. From this figure, it is clear that our LSB sample peaks at around $g - r \sim 0.4$ which is on the blue side. Indeed, if we use $(g-r)_{\mathrm{cut}} =0.65-0.03\,(M_r+20)$ \citep{blanton+05} to divide our LSB sample into red and blue populations, the dominant population of our sample galaxies is blue. The existence of these blue LSBs suggests that they are not just the faded remnants of the HSBs. When compared to the SDSS all-galaxy sample, these LSB galaxies occupy the bluer region of the colour-magnitude diagram. It remains to determine what fraction of our sample LSB galaxies is faint. A natural boundary between the bright and faint galaxies is the absolute magnitude ($M_{*}$) corresponding to an $L_{*}$ galaxy, appearing in the Schechter luminosity function \citep{blanton+03}. We classify a LSB galaxy as faint if its absolute magnitude is larger than $M_{*} + 2$ \citep{Hollosi-Efstathiou1988}. Based on the SDSS $r$-band local luminosity function, we have $M_{*}= - 20.66$ \citep{blanton+03} or $-20.71$ according to the GAMA survey \citep{Lovedayetal2015}. According to this criteria, about 71 $\%$ of our sample LSB galaxies are on the faint side; some are as faint as $M_{r} \sim -14$ and can definitely be classified as ultra-faint.  Figure~\ref{fig:colour_diskmag} shows the location of our barred and unbarred LSBs on the CMR. Most of the barred and unbarred LSBs are blue in our sample and shows no preference for the colour. However, \cite{masters+11} has found an increase in bar fraction in redder galaxies.

Figure~\ref{fig:bar_lum} shows the bar fraction of LSB galaxies as a function of $r$- band absolute magnitude. The barred galaxies are systematically brighter than the unbarred galaxies. This implies that there is higher probability of finding bars in luminous galaxies.
There are  also 7 red LSBs in our sample and 1 of them has bar in it. In the section below, we will return to the case of barred LSB population.
  
\subsection{Correlation between colour, stellar mass and central surface brightness}
\label{sec:color}
In Figure~\ref{fig:struc_lsb}, we present the correlation between various structural properties such as central surface brightness, colours and stellar mass of galaxies. Stellar masses are calculated following the \cite{bell+03} prescription. In that, we compute the $r$-band mass-to-light ratios $(M/L)_r$ for our sample galaxies using the $g - r$ colour.  The left panel shows the $\mu_{0,\rm{disk}}$ vs $g - r$ colour of the sample LSB galaxies. As mentioned above, most of our LSB galaxies are blue, but with a wide variation in their central surface brightness. Further, the central surface brightness shows no correlation with the host galaxy stellar mass (see the right panel). The middle panel shows the scatter plot of the stellar mass and the $g-r$ colour for our LSB galaxies.  As it seems, there is a slight tendency that high stellar mass LSB galaxies, on an average, are redder in colour.

From the middle panel of Figure~\ref{fig:struc_lsb}, we see a weak trend in the incidence of a bar and $g - r$ colour and the host galaxy stellar mass. Bars seem to be associated with higher stellar mass and $g-r$ colour, although most galaxies in our sample are bluer. Such a trend amongst our LSB galaxies is in compliance with late type spirals \citep[see][]{Barazzaetal2008, masters+11}. The faintest barred LSB galaxy in  our sample has disk central surface brightness $\sim 23.28$~mag/arcsec$^2$ and its colour is on the bluer side.  

Further, we find no correlation on the incidence of a bar with the host galaxy central surface brightness. As bars are known to exist in galaxies with little or no classical bulge (e.g., our MW), the incidence of a bar may not entirely be decided by the pre-existing bulge. But when HSB and LSB galaxies are compared, bars are seen more prominently in HSB galaxies rather than in LSB galaxies \citep[see ]{masters+11}. However, in our LSB sample for which the disk central surface brightness varies from 24 - 21 mag/arcsec$^2$, we do not find any preference regarding the incidence of a bar.
 
\section{Barred LSB galaxies vs HI gas mass} 
\label{sec:bar-HIgas}

\begin{figure}
\includegraphics[width=0.45\textwidth]{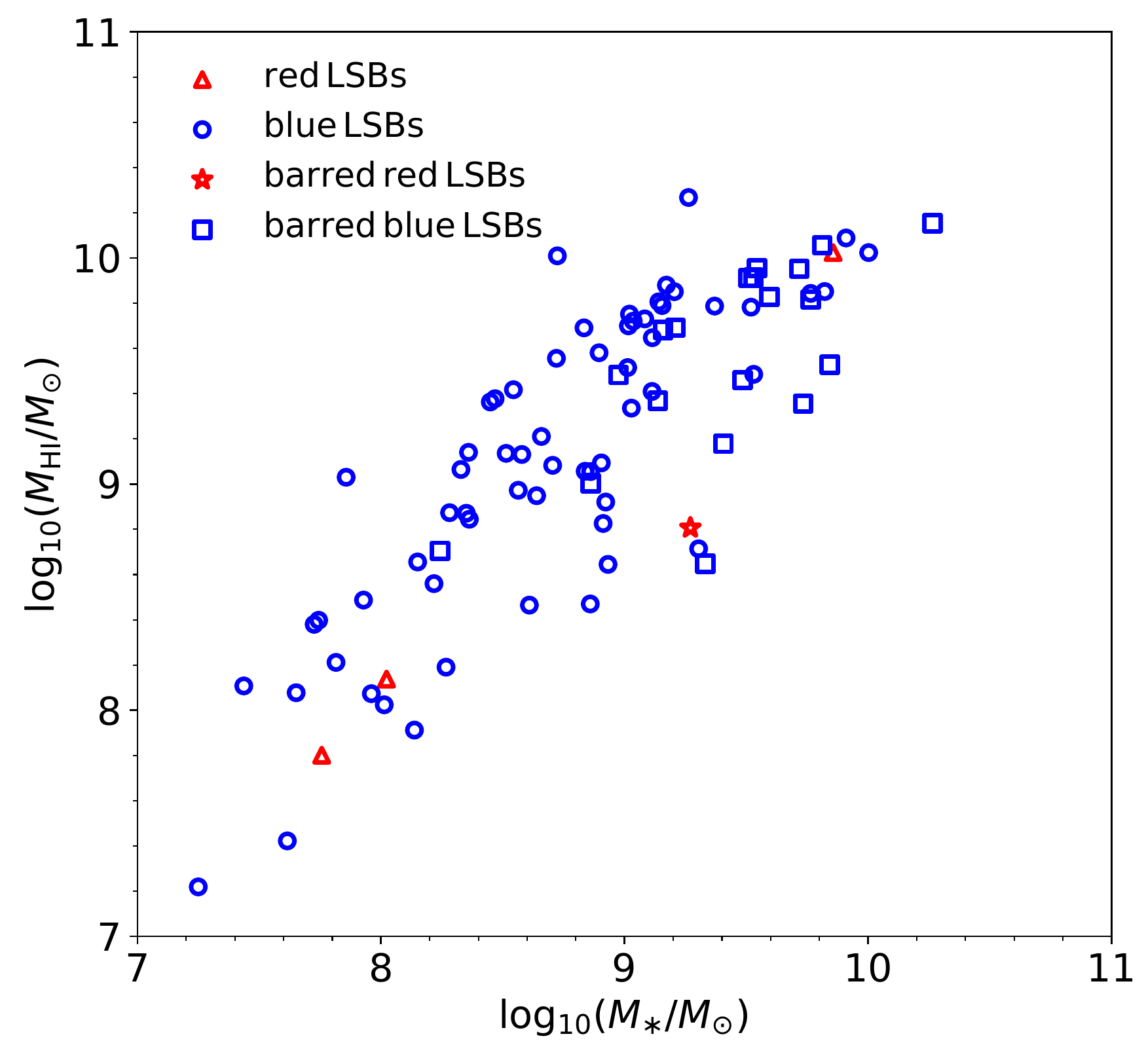}
\caption{Stellar mass versus HI mass of the LSB galaxies in our sample. The blue and red LSB galaxies are shown as blue circles and red triangles whereas the barred red and blue galaxies are marked by red stars and blue squares.}
\label{fig:HI_mgal}
\end{figure}

\begin{figure}
\includegraphics[width=0.45\textwidth]{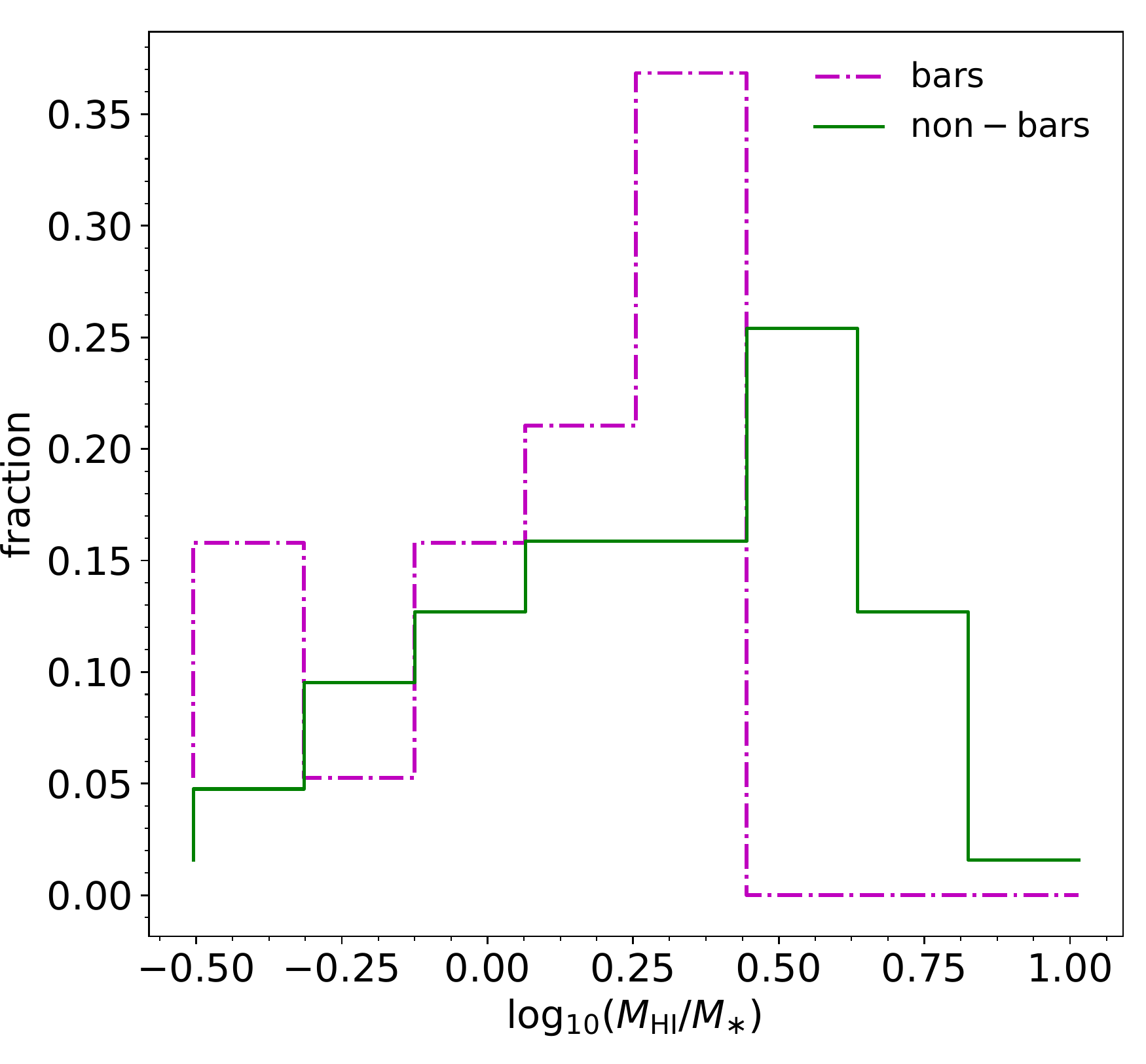}
\caption{Histogram of HI mass to stellar mass of the LSB galaxies in our sample. Most of the galaxies in our sample are gas-rich.}
\label{fig:HIbymgal}
\end{figure}

The presence of gas plays a major role in the formation of bars in disk galaxies. Numerical simulations have demonstrated that as the gas fraction increases, the bars get weaker and when the fraction is close to unity or more, simulations end up with almost no bar \citep{Athanassoulaetal2013}. Earlier studies have shown that dissipative effect of gas might even lead to the bar destruction \citep{BournaudCombes2002}. Keeping this in mind, we investigate the incidence of bars in LSB galaxies in the presence of cold neutral hydrogen gas measured by the Green Bank Telescope \citep{schneider+92}. Only 85 galaxies in our LSB sample have reliable HI observations and these are shown in Figure~\ref{fig:HI_mgal}. It is clear that there is a strong correlation between the stellar mass and HI gas mass for our LSB galaxies. Most of these LSB galaxies are very gas-rich,  with gas fraction $f_{gas} = M_{HI}/M_{*} > 1$ (Figure~\ref{fig:HIbymgal}). Of the 85 LSBs, there are $4$ red LSBs and the rest are blue. {\it We see bars both in red and blue LSB galaxies. Irrespective of their colours, the incidence of a bar is associated with high gas fraction in our sample.}  Not only that, we have about $37$ blue LSBs hosting bars - this implies that about $15\%$ of the blue LSB galaxies in our sample host bars  - this number is in sync with other recent studies \citep{masters+12, SG17} based on large volume limited sample drawn from SDSS DR7. These studies and a number of others have shown that this fraction is lower than that in gas-poor spirals \citep{Eskridge+2000, Barazzaetal2008,masters+11}. However, according to a recent study \citep{erwin18}, bars are common in both gas-rich blue galaxies as well as in gas-poor red spirals. It remains to be understood what makes bar formation possible in such gas-rich (with $f_{gas} \gtrsim 1$) blue LSB galaxies.

\section{Discussion and Conclusions}
\label{sec:discussion}
We have studied a sample of 263 LSB galaxies observed by Green Bank Telescope \citep{schneider+92} which are in overlap with the SDSS footprint. We have performed two-component bulge-disk decomposition of 263 galaxies in the SDSS $g$, $r$ and $i$ bands and investigated their structural properties in detail. We have found that $60\%$ LSBs in our specific sample are bulgeless while $40\%$ are with bulges. Some of the LSBs are associated with significant bulge component with $B/T > 0.1$. Since LSBs are known to be dwelling in less dense environment \citep{RosenbaumBomans2004}, mergers and interactions are unlikely to have led the bulge formation. We also have $15 \% $ barred galaxies in our sample. Our findings of bulges and bars suggest a considerable on-going evolution  in the local LSB galaxies and the bars might as well be playing a role in the bulge growth \citep{laur+07,gadotti11,cheung+13}. The interesting fact about our sample is that they are not the class of giant LSB galaxies, in fact, most of our LSBs are faint, blue and gas-rich and roughly half of them are hosting bars and bulges. Since LSB galaxies are dark matter dominated, disks are known to be stable against bar formation \citep{OstrikerPeebles1973,efsth+82,chris+95,Sodietal2015,algorry+17}, as shown in numerical simulations of stellar disks with dark matter dominance at all radii \citep{Saha2014}. Question arises how these faint blue LSBs are making bars and bulges. One possibility is that LSB disks are embedded in dark matter halos that are spinning \citep{jimenez+98,Vitvitskaetal2002,KL13} which might be promoting bar formation provided spin parameter is not too high \citep{SahaNaab2013,sodi+13,long+14,collier+17}. Whether these bars lead to the formation of bulges or other processes such as minor mergers being involved, needs further and detailed investigation. 

Major conclusions from our work on this {\it specific} sample are:

\begin{itemize}

\item We classify a galaxy as LSB if its $r$-band disk central surface brightness is fainter than 21 $\mathrm{mag \, arcsec^{-2}}$. According to the threshold criterion, $\sim$84\% galaxies in our sample are LSB. This fraction reduces further to 61\% when the central surface brightness of the full galaxy light distribution is used.

\item The median scale-length of LSB sample galaxies is 2.4 kpc which is comparable to the full galaxy population (average $\sim$3.79 kpc) as shown by \cite{fathi+10}. We have found no correlation between the scale length and disk central surface brightness. 

\item There seems to be a weak correlation between the colour and stellar mass of these LSB galaxies.

\item Dominant fraction of galaxies in our sample is faint, with absolute magnitude as faint as $-14$. This sample of faint LSBs is rich in morphology.

\item Most of our LSB galaxies are blue as per $g-r$ colour criteria. However, there are also 7 red LSBs in our sample.

\item Based on the bulge-disk decomposition, we have found that $40\%$ of our sample LSBs are with bulges and of these, there are $\sim 20\%$ with $B/T > 0.1$. 

\item We have found that $\sim 15\%$ LSBs in our sample are barred. Bars are seen in both red and blue LSBs in our sample. The incidence of a bar has no correlation on the host galaxy central surface brightness.  Most of these barred LSBs are highly gas-rich and blue. 

\end{itemize}

\section*{Acknowledgments}
\noindent
We thank the anonymous referee for a careful reading and insightful comments on the manuscript.
The research of IP is supported by the INSPIRE Faculty grant (DST/INSPIRE/04/2016/000404) awarded by the Department of Science and Technology, Government of India.

SDSS is managed by the Astrophysical Research Consortium for the Participating Institutions of the SDSS Collaboration including the Brazilian Participation Group, the Carnegie Institution for Science, Carnegie Mellon University, the Chilean Participation Group, the French Participation Group, Harvard-Smithsonian Center for Astrophysics, Instituto de Astrofísica de Canarias, The Johns Hopkins University, Kavli Institute for the Physics and Mathematics of the Universe (IPMU) / University of Tokyo, Lawrence Berkeley National Laboratory, Leibniz Institut für Astrophysik Potsdam (AIP), Max-Planck-Institut für Astronomie (MPIA Heidelberg), Max-Planck-Institut für Astrophysik (MPA Garching), Max-Planck-Institut für Extraterrestrische Physik (MPE), National Astronomical Observatories of China, New Mexico State University, New York University, University of Notre Dame, Observatório Nacional / MCTI, The Ohio State University, Pennsylvania State University, Shanghai Astronomical Observatory, United Kingdom Participation Group, Universidad Nacional Autónoma de México, University of Arizona, University of colourado Boulder, University of Oxford, University of Portsmouth, University of Utah, University of Virginia, University of Washington, University of Wisconsin, Vanderbilt University, and Yale University.

We have also used PyRAF for our study, a product of the Space Telescope Science Institute, which is operated by AURA for NASA.

\bibliography{LSB}
\appendix

\section{Comparison of model magnitudes with the SDSS Petrosian magnitudes}
\begin{figure*}
\centering
\includegraphics[width=0.8\textwidth,height=0.4\textwidth]{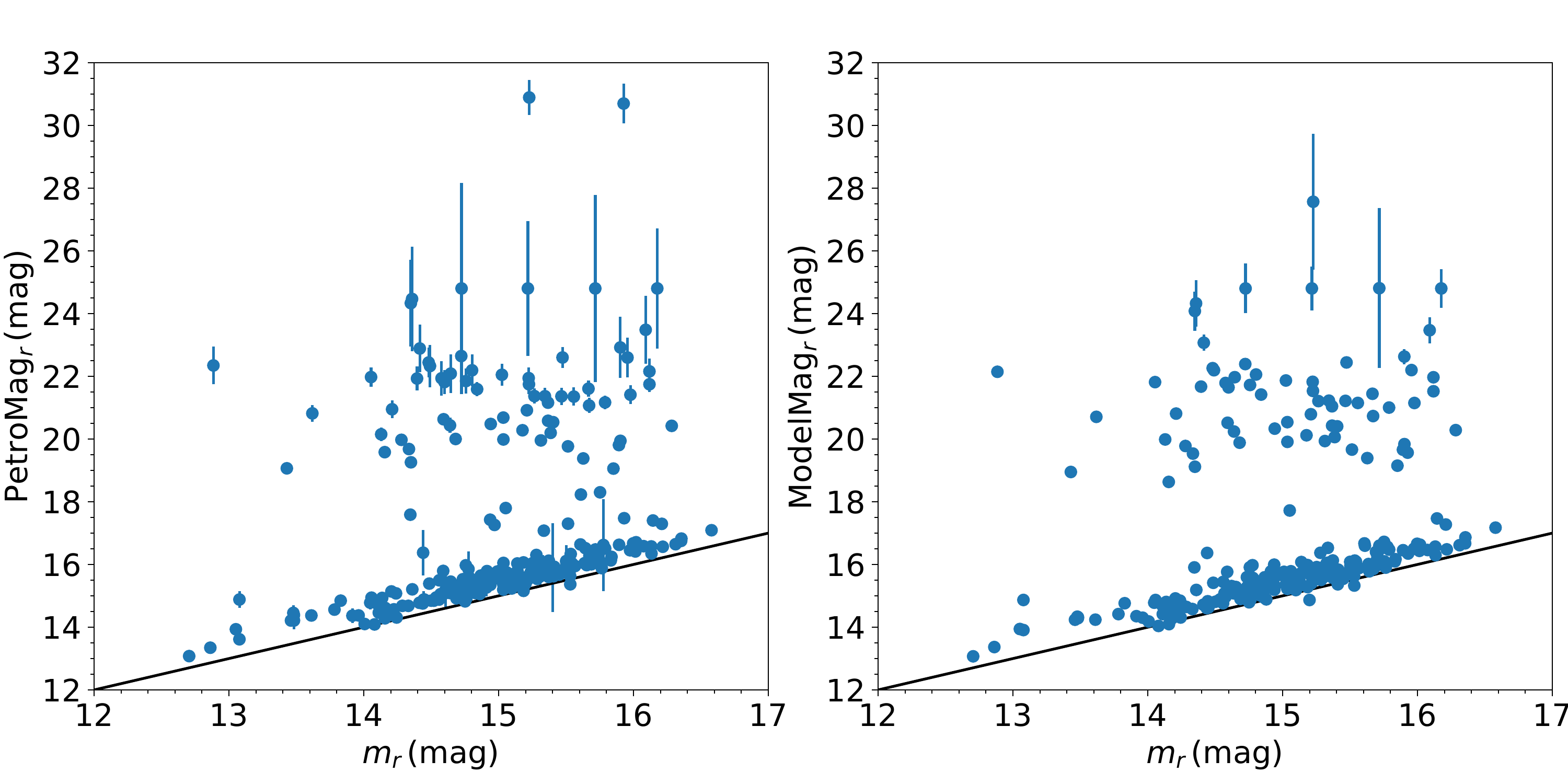}
\caption{Comparison of SDSS Petrosian (left panel) and model (right panel) magnitudes with the AB magnitudes of LSB sample galaxies in $r$-band. The AB magnitudes of the sample galaxies are calculated from the flux estimated by the model magnitudes given by GALFIT. The black colour solid line in both panels is 45$^\circ$ line for a better visual comparison. 
\label{fig:mag_comp}}
\end{figure*} 
In this appendix, we give the comparison of magnitudes of the LSB sample galaxies as given by the SDSS with that of what we get from GALFIT decomposition. We convert the $r$-band GALFIT model magnitudes of LSB galaxies to the AB magnitudes using standard relation \citep{Oke1974}. In the left hand panel of Figure~\ref{fig:mag_comp}, we show the comparison of SDSS Petrosian magnitudes and the AB magnitudes of LSB galaxies  in $r$-band whereas the right hand panel depicts the comparison of  SDSS model magnitudes with that of AB magnitudes. Galaxies which have $r$-band Petrosian magnitudes less than 17, are pretty much in agreement with our AB magnitudes of galaxies. However, almost all the galaxies have bit brighter AB magnitudes as compared to their SDSS Petrosian or model magnitudes. There are around $\sim$ 28\% galaxies where the difference between SDSS Petrosian or model magnitudes and AB magnitudes is more than 1 mag.  If we consider Petrosian or model magnitudes from SDSS for our analysis, more galaxies would be termed as LSB.

\label{lastpage}

\end{document}